\documentclass{PoS}
\usepackage{cancel}

\title{Supersymmetric signals in $Z'$ decays}

\ShortTitle{Supersymmetric signals in $Z'$ decays}

\author{\speaker{Gennaro Corcella}\\
  INFN, Laboratori Nazionali di Frascati\\
  Via E.~Fermi 40, 00044 Frascati (RM), Italy\\
        E-mail: \email{gennaro.corcella@lnf.infn.it}}

\abstract{
  I present a scenario wherein heavy neutral vector bosons $Z'$,
  predicted by GUT-inspired U(1)$'$ models, 
  decay into supersymmetric final states, besides the
  Standard Model channels investigated at the LHC.
  It is found that accounting for such 
  decays lowers the exclusion limits on the $Z'$ mass at 13 TeV
  by about 200-300 GeV.
}

\FullConference{XXV International Workshop on Deep-Inelastic
  Scattering and Related Subjects\\
		3-7 April 2017\\
		University of Birmingham, UK}

\begin{document}

\section{Introduction}
Heavy neutral gauge bosons $Z'$ are predicted by extensions of
the Standard Model (SM) based on U(1)$'$ gauge symmetries, typically inspired
by Grand Unification Theories (GUT) \cite{rizzo,langa}.
They are also present in the so-called
Sequential Standard Model (SSM), the simplest extension of the
SM, wherein $Z'$ and possibly $W'$ bosons have the same coupling
to fermions as $Z$ and $W$.

The LHC experiments have searched for the $Z'$, assuming that
it decays only into Standard Model channels and focusing on high-mass
charged-lepton pairs or dijets.
As for dileptons, 
by using the 13 TeV LHC data, 
the ATLAS Collaboration \cite{atlasll} set the limits 
$m_{Z'}>4.5$~TeV in the SSM and $m_{Z'}>3.8$-4.1~TeV in
U(1)$'$ models, whereas CMS obtained $m_{Z'}>4.0$~TeV (SSM) and   
$m_{Z'}>3.5$~TeV (GUT-inspired models) \cite{cmsll}.
For dijets, the limits read $m_{Z'}>2.1$-2.9~TeV (ATLAS) \cite{atlasjj}
and $m_{Z'}>2.7$~TeV (CMS) \cite{cmsjj}. 
In this talk, I explore possible supersymmetric decays of 
GUT-inspired $Z'$ bosons, within the Minimal Supersymmetric
Standard Model (MSSM), along the lines of 
Refs.~\cite{corgen,cor}.
In fact, the opening of new channels lowers the 
SM branching ratios and the exclusion
limits; moreover, if a $Z'$ were to be discovered,
decay modes like $Z'\to \tilde\ell^+\tilde\ell^-$ would be an ideal
place to search for supersymmetry, since the $Z'$ mass
would set a constraint on sparticle invariant masses.
Also, decays of the $Z'$ into the lightest supersymmetric particles,
such as neutralinos, would be useful from the viewpoint of
Dark Matter searches at colliders.

I shall first discuss the main theoretical features of the considered
model and then present some results for a representative benchmark point
of the parameter space.
I will finally make some concluding remarks.

\section{Theoretical framework}
U(1)$'$ gauge groups and $Z'$ bosons typically arise from the breaking of a 
rank-6 Grand Unification group ${\rm E}_6$.
The $Z'_\psi$ is associated with U(1)$'_\psi$, coming from the
breaking of E$_6$ into SO(10):
\begin{equation}\label{upsi}
{\rm E}_6\to {\rm SO}(10)\times {\rm U}(1)'_\psi.
\end{equation}
The subsequent breaking of SO(10) leads to the 
$Z'_\chi$ boson:
\begin{equation}
{\rm SO}(10)\to {\rm SU}(5)\times {\rm U}(1)'_\chi.
\end{equation}
A generic $Z'$ is then given by the mixing between
$Z'_\psi$ and $Z'_\chi$ bosons via an angle $\theta$:
\begin{equation}\label{ztheta}
Z'(\theta)=Z'_\psi\cos\theta-Z'_\chi\sin\theta.
\end{equation}
Another scenario, typical of superstring theories, 
consists in the breaking of E$_6$ in the SM and a U(1)$'_\eta$ group, leading
to a $Z'_\eta$ boson, with a mixing angle 
$\theta=\arccos\sqrt{5/8}$:
\begin{equation}\label{ueta}
{\rm E}_6\to {\rm SM}\times U(1)'_\eta.
\end{equation}
This talk will be concentrated on the
the $Z'_\psi$ model, as it is the most promising, yielding the
largest cross section.
Scenarios with other values of the angle $\theta$ 
were discussed in detail in
\cite{corgen,cor}.

As for the MSSM, 
besides the scalar Higgs doublets $H_d$
and $H_u$, an extra neutral singlet $S$ is necessary
to break the U(1)$'$ gauge symmetry and give mass to the $Z'$.
After electroweak symmetry breaking, the Higgs sector consists of
one pseudoscalar $A$ and three scalars $h$, $H$ and $H'$, where
$H'$ is a new boson, due to the extra U(1)$'$ symmetry.
In the gaugino sector,  
two extra neutralinos are present, associated with the
supersymmetric partners of $Z'$ and $H'$, while the charginos
are unchanged.
As thoroughly debated in \cite{gherghetta}, the U(1)$'$ group
leads to extra D- and F-term
contributions to sfermion masses, which were accounted for in
\cite{corgen,cor} and will be included in the results.
In the following section,
I shall explore the U(1)$'_\psi$ model,
which exhibits the most interesting phenomenology
with $Z'$ decays into supersymmetric final states.

\section{Results}
The $Z'_\psi$ model corresponds to a mixing angle $\theta=0$;
in the present analysis, the $Z'$ mass
will be set to 
$m_{Z'}=2~{\rm TeV}$ and the coupling
constants of U(1)$'$ and U(1)$_{\rm Y}$ 
proportional according to $g'=\sqrt{5/3}~g_1$.
I shall also set:
$M_1=400$~GeV, $M'=1$~TeV, $\tan\beta$=30,
$\mu=200$~GeV and
$A_q=A_\ell=A_\lambda=4$~TeV, where $A_q$ and $A_\ell$ are the soft
trilinear couplings of squarks and sleptons with the Higgs fields
and $A_\lambda$ is the soft Higgs trilinear coupling.

The sfermion spectrum, corresponding to soft masses at the $Z'$ scale
$m_{\tilde\ell}^0=m_{\tilde\nu_\ell}^0=1.2$~TeV and 
$m^0_{\tilde q}$=5.5~TeV, computed by means of the SARAH \cite{sarah}
and SPheno \cite{spheno} codes,
is given in Tables 
\ref{tabmassq} and \ref{tabmassl}, where $\tilde\ell_{1,2}$ and
$\tilde q_{1,2}$ denote the mass eigenstates.
Tables~\ref{tabmassh} and \ref{tabmasscn} contain instead the mass spectra
of Higgs bosons and gauginos (charginos and neutralinos), respectively.
  \begin{table}[htp]
\caption{Squark masses in GeV in the reference point of the MSSM.}
\label{tabmassq}
\begin{center}
\small
\begin{tabular}{|c|c|c|c|c|c|}
\hline
$m_{\tilde d_1}$ &   $m_{\tilde u_1}$ &  $m_{\tilde s_1}$ &  $m_{\tilde c_1}$ &  
$m_{\tilde b_1}$ & $m_{\tilde t_1}$\\ 
\hline
 5609.8 & 5609.4  & 5609.9 & 5609.5 & 2321.7 & 2397.2 \\
\hline
$m_{\tilde d_2}$ &   $m_{\tilde u_2}$ &  $m_{\tilde s_2}$ &  $m_{\tilde c_2}$ &  
$m_{\tilde b_2}$ & $m_{\tilde t_2}$\\ 
\hline
 5504.9 & 5508.7  & 5504.9 & 5508.7 & 2119.6 & 2036.3 \\
\hline\end{tabular}
\end{center}
\end{table}
\begin{table}[htp]
\caption{Mass spectrum in GeV of charged sleptons ($\ell=e,\mu$)
and sneutrinos.}
\label{tabmassl}
\begin{center}
\small
\begin{tabular}{|c|c|c|c|c|c|c|c|}
\hline
$m_{\tilde \ell_1}$ &   $m_{\tilde \ell_2}$ &  $m_{\tilde\tau_1}$ & $m_{\tilde\tau_2}$ &
$m_{\tilde \nu_{\ell,1}}$ &  $m_{\tilde \nu_{\ell,2}}$ &
$m_{\tilde \nu_{\tau,1}}$ &  $m_{\tilde \nu_{\tau,2}}$ \\ 
\hline
 1392.4 & 953.0  & 1398.9 & 971.1 & 1389.8  & 961.5 & 1395.9 & 961.5\\ 
\hline\end{tabular}
\end{center}
\end{table}
\begin{table}[htp]
\caption{Masses of neutral and charged Higgs bosons in GeV.}
\label{tabmassh}
\begin{center}
\small
\begin{tabular}{|c|c|c|c|c|}
\hline
$m_h$ &   $m_H$&  $m_{H'}$ &  $m_A$ & $m_{H^\pm}$\\ 
\hline
 125.0 &  1989.7 & 4225.0  & 4225.0 & 4335.6 \\ 
\hline\end{tabular}
\end{center}
\end{table}
\begin{table}[htp]
\caption{Masses of charginos 
and neutralinos in GeV.}
\label{tabmasscn}
\begin{center}
\small
\begin{tabular}{|c|c|c|c|c|c|c|c|}
\hline
$m_{\tilde\chi_1^+}$ &   $m_{\tilde\chi_2^+}$ & $m_{\tilde\chi_1^0}$ &   $m_{\tilde\chi_2^0}$ 
& $m_{\tilde\chi_3^0}$ &   $m_{\tilde\chi_4^0}$ & $m_{\tilde\chi_5^0}$ &   $m_{\tilde\chi_6^0}$  \\ 
\hline
 204.8 & 889.1 & 197.2  & 210.7 & 408.8 & 647.9 & 889.0 & 6193.5 \\ 
\hline\end{tabular}
\end{center}
\end{table}\par
In this benchmark point, 
the $Z'_\psi$ branching ratios were computed in \cite{cor}:
although the SM modes are still dominant, 
the overall branching ratio into supersymmetric final states is almost
30\%. The highest supersymmetric rate is the one into chargino pairs 
$\tilde\chi^+_1\tilde\chi^-_1$ and accounts
for about 10\%;
the branching ratio into the 
lightest neutralinos, Dark Matter candidates, is
roughly 5\%.

As in \cite{cor}, one can consider the decay chain
$pp\to Z'_\psi\to\tilde\chi_1^+\tilde\chi_1^-\to
(\tilde\chi_1^0\ell^+\nu_\ell)(\tilde\chi_1^0\ell^-\bar\nu_\ell)$,
with $\ell=\mu, e$, leading to final
states with two charged leptons and missing energy.
Its cross section, calculated at leading order (LO) 
by MadGraph \cite{madgraph}, is given by
$7.9\times 10^{-4}$~pb
at 14 TeV.
Hereafter, I will present some relevant leptonic distributions
and compare them with decays $Z'_\psi\to \ell^+\ell^-$
and direct chargino production, namely
$pp\to \tilde\chi_1^+\tilde\chi_1^-\to
(\tilde\chi_1^0\ell^+\nu_\ell)(\tilde\chi_1^0\ell^-\bar\nu_\ell)$.
The hard-scattering processes will be simulated by MadGraph and 
matched with HERWIG 6 \cite{herwig} for shower and hadronization.

\begin{figure}[htp]
\centerline{\resizebox{0.45\textwidth}{!}
{\includegraphics{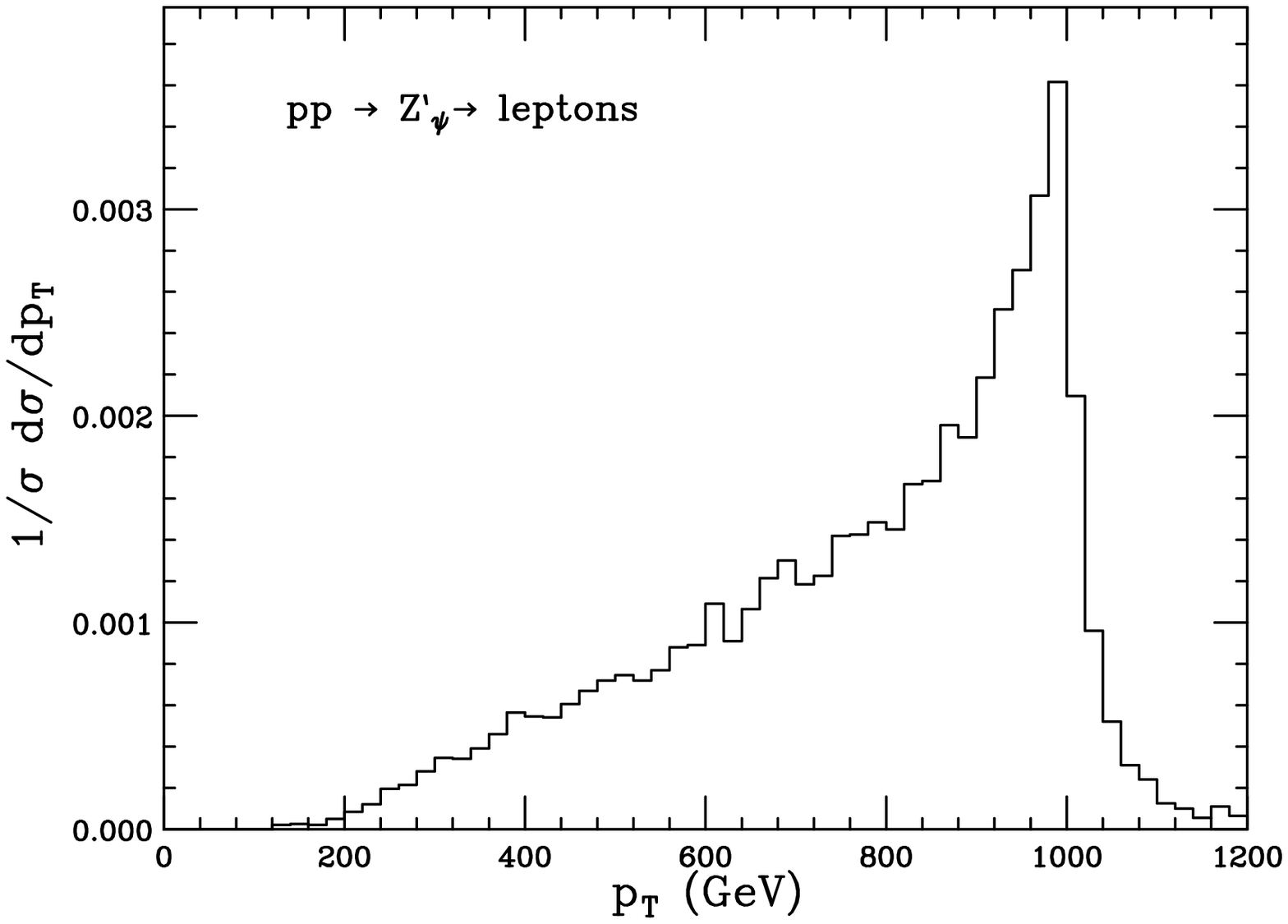}}%
\hfill%
\resizebox{0.45\textwidth}{!}{\includegraphics{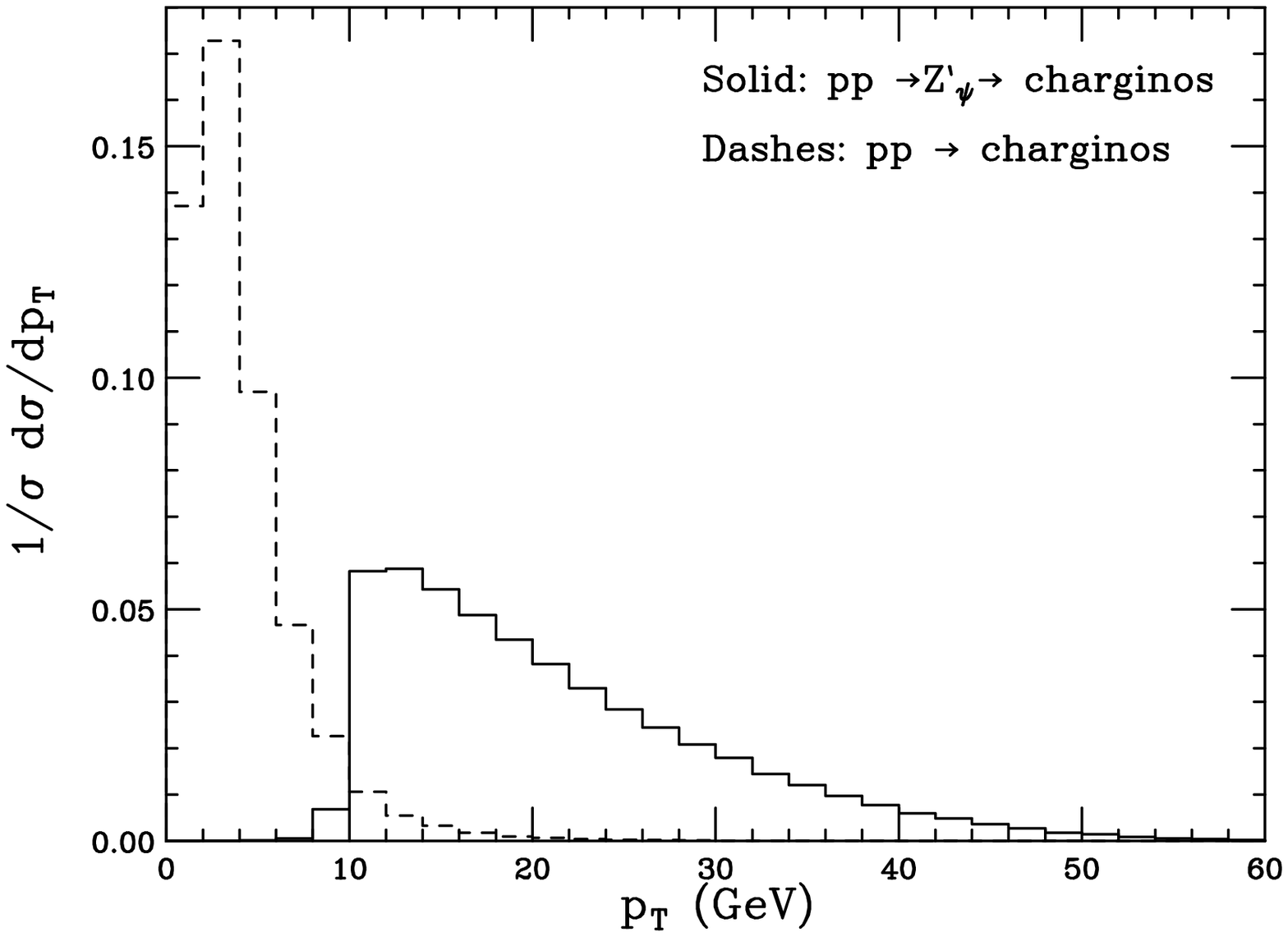}}}
\caption{Lepton transverse momentum for the $Z'_\psi$ model at $\sqrt{s}=14$~TeV
and $m_{Z'}=2$~TeV, for a direct $Z'_\psi\to \ell^+\ell^-$ decay (left) and 
chains initiated by $Z'_\psi\to\tilde\chi^+_1\chi^-_1$ or direct 
chargino production processes (right).}
\label{zpsipt}
\end{figure}

Figure~\ref{zpsipt} presents the transverse-momentum spectrum of the leptons
produced in all three processes.
For direct $Z'_\psi\to\ell^+\ell^-$, 
the two leptons get the full initial-state transverse momentum
and the $p_T$ spectrum is relevant at high values;
in the other cases, 
some (missing) transverse momentum is lent to
neutrinos and neutralinos, which significantly decreases the $p_T$ of $\ell^+$
and $\ell^-$.  
For direct chargino production,
the leptons are rather soft and the spectrum is peaked at low $p_T$;
when the charginos come from the
$Z'$, their invariant mass must be equal to $m_{Z'}$ and therefore the
transverse-momentum distribution is substantial at higher $p_T$ values.
\begin{figure}
\centerline{\resizebox{0.45\textwidth}{!}{\includegraphics{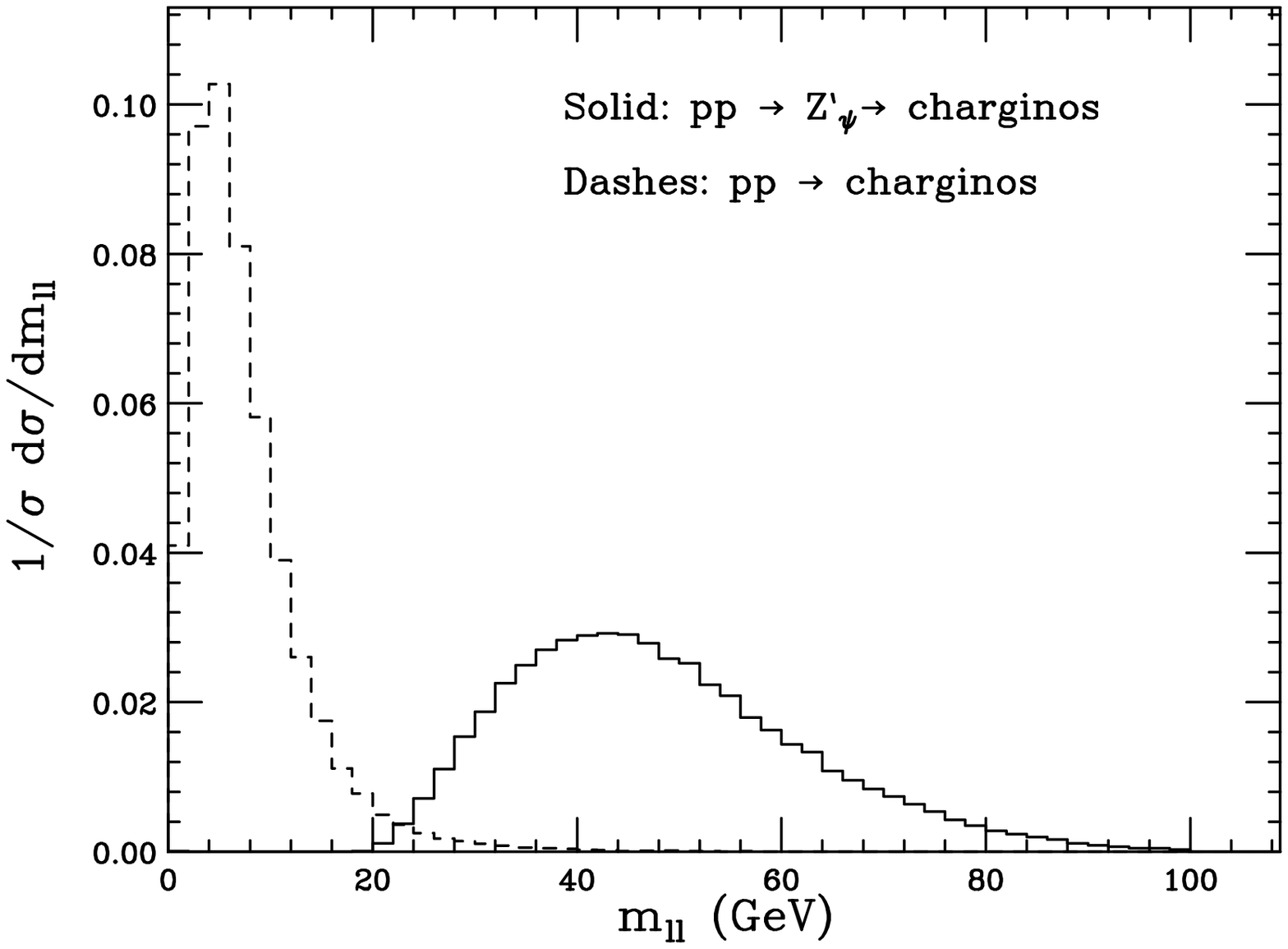}}%
\hfill%
\resizebox{0.45\textwidth}{!}{\includegraphics{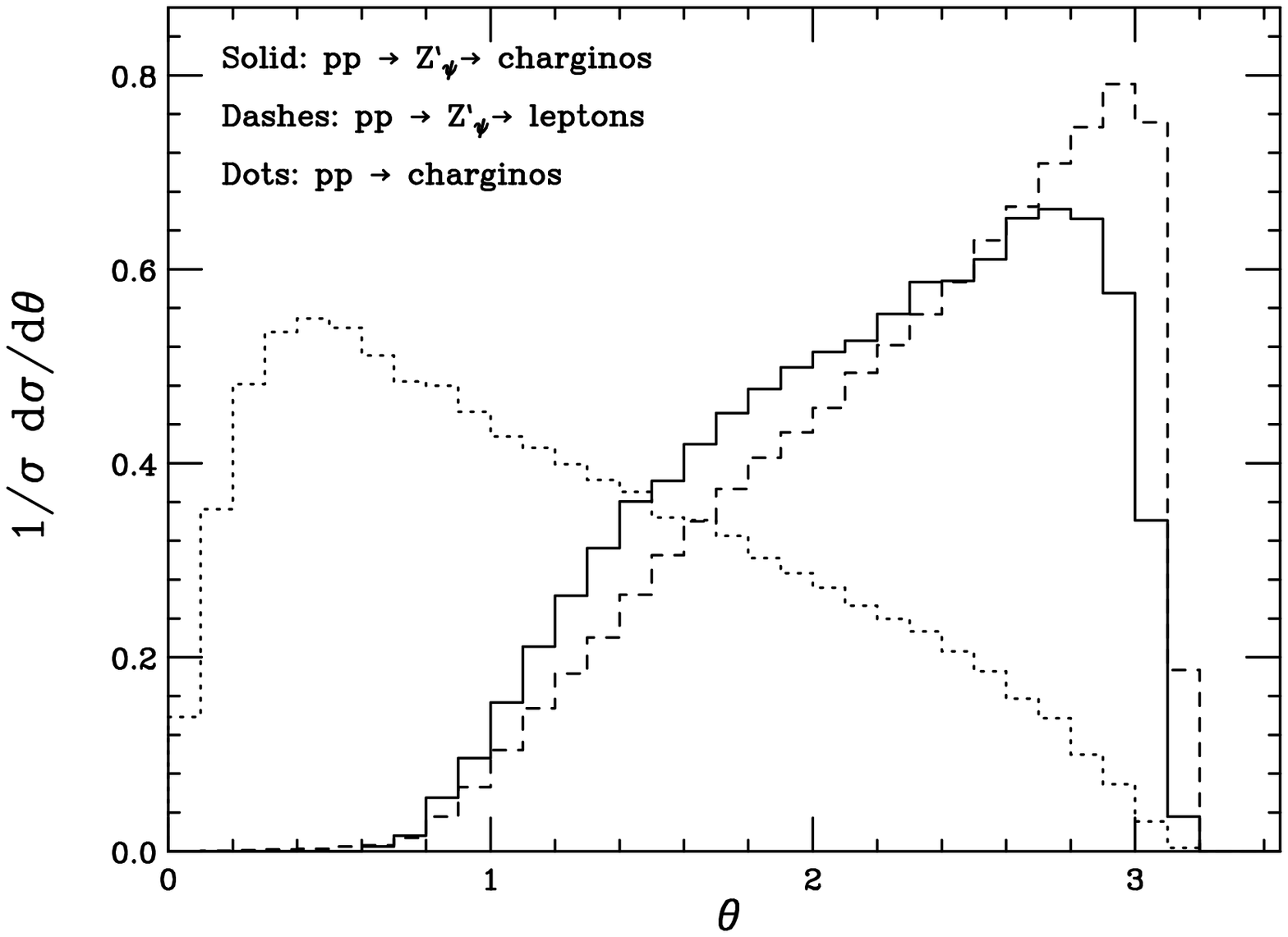}}}
\caption{Left: dilepton invariant mass, with $\ell^+$ and 
$\ell^-$ 
originated from $Z'_\psi$ decays into charginos and for
direct $pp\to \tilde\chi^+_1\tilde\chi^-_1$ production.
Right: angle between $\ell^\pm$ in the laboratory frame, in all three
processes.}
\label{zpsimth}
\end{figure}
\par
Figure~\ref{zpsimth} presents the dilepton invariant mass
$m_{\ell\ell}$ (left), as well as
the angle $\theta$ between $\ell^+$ and $\ell^-$ (right).
For $Z'$ decays into charginos, $m_{\ell\ell}$ 
lies in the range 20 GeV~$<m_{\ell\ell}<$~100 GeV and is
maximum at $m_{\ell\ell}\simeq 45$~GeV;
for direct $\tilde\chi_1^+\tilde\chi_1^-$ production, $m_{\ell\ell}$ 
is peaked about 5 GeV and rapidly decreases.
As for the $\theta$ spectrum, in $Z'_\psi$ direct leptonic decays
it exhibits a maximum 
about $\theta\simeq 3$, i.e. $\ell^+$ and $\ell^-$ almost
back-to-back;
in $Z'_\psi\to\tilde\chi^+_1\tilde\chi^-_1$, 
the $\theta$ distribution is broader and peaked
at a lower $\theta\simeq 2.75$;
for direct chargino-pair production,  
$\ell^+$ and $\ell^-$ are mostly soft and collinear and 
the angular distribution is peaked and substantial
at smaller $\theta$.

\begin{figure} 
\centerline{\resizebox{0.45\textwidth}{!}{\includegraphics{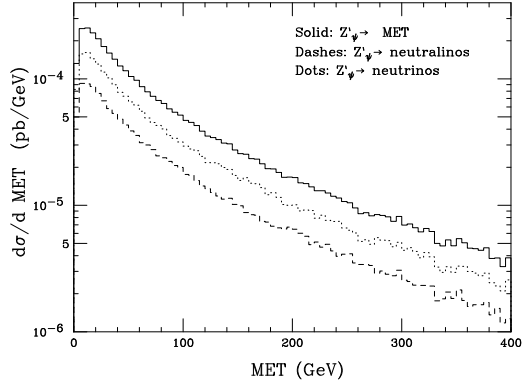}}}
\caption{MET in $ Z'_\psi$ decays: plotted are 
  the neutralino (dashes), neutrino (dots) and total (solid)
rates.}
\label{metneu}
\end{figure}

Decays into neutralino pairs $\tilde\chi^0_1\tilde\chi^0_1$
are relevant for the searches
for Dark Matter candidates and exhibit a cross section
$\sigma(pp\to Z'_\psi\to \tilde\chi^0_1\tilde\chi^0_1)\simeq 
6.4\times 10^{-3}$~pb at 14 TeV;
competing processes are $Z'_\psi$ decays into neutrino pairs.
Figure~\ref{metneu} displays the total
missing transverse energy (MET) spectrum and the contribution due to the 
neutrino and neutralino pairs; all spectra
are normalized to the LO cross section.
The shapes of all distributions are roughly the same, but
the total number of events
at any MET is substantially higher, by about 60\%,
if neutralinos contribute.
\begin{figure} 
\centerline{\resizebox{0.45\textwidth}{!}{\includegraphics{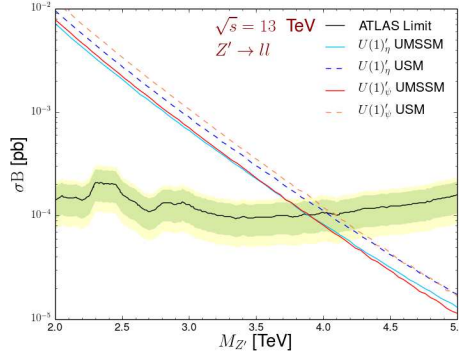}}}
\caption{Comparison of ATLAS high-mass dilepton
  data at 13 TeV with the theoretical
  predictions for $\sigma(pp\to Z'){\rm BR}(Z'\to\ell^+\ell^-)$
  in U(1)$'_\psi$ and U(1)$'_\eta$ models, with (UMSSM) or without (USM)
supersymmetry.}
\label{atlas}
\end{figure}
\par Before completing this section, following \cite{araz},
in Fig.~\ref{atlas}
I compare
the ATLAS data on high-mass dileptons at 13 TeV
with the predictions
for the $Z'_\psi$ and $Z'_\eta$ models, possibly accounting
for supersymmetric decays, expressed in terms of
the product $\sigma(pp\to Z')~{\rm BR}(Z'\to\ell^+\ell^-)$.
From the comparison, one learns that the impact of the 
supersymmetric modes on the
exclusion limits on the mass of $Z'_\eta$ and $Z'_\psi$ bosons
is about 200 and 300 GeV, respectively.

\section{Conclusions}
I investigated supersymmetric $Z'$ decays at the LHC,
for $\sqrt{s}=14$~TeV in the U(1)$'_\psi$ model.
In this scenario, $Z'$ decays into
into chargino pairs lead to final
states with leptons and missing energy which
can be discriminated from 
$Z'\to\ell^+\ell^-$ processes as well as from direct
chargino production.
Decays into the lightest neutralinos 
substantially contribute to the missing transverse energy
and can therefore be explored in Dark Matter searches.
Comparing with the ATLAS data at 13 TeV, one learns that  
the inclusion of supersymmetric modes has an impact
of the order of 200-300 GeV on the $Z'$ mass limits.

In a future perspective,
a complete analysis should necessarily compare supersymmetric
signals in $Z'$ decays with the backgrounds coming from the 
SM and other supersymmetric processes, 
as well as non-supersymmetric $Z'$ decays,
and implement the detector simulation. 
Furthermore, in order to further enhance
non-standard signals in $Z'$ events, one may consider $Z'$ 
supersymmetric decays in the framework of leptophobic models, wherein $Z'$
decays into electron or muon pairs are suppressed \cite{araz}.


\begin{thebibliography}{99}
\bibitem{rizzo}
J.L. Hewett and T. Rizzo, Phys. Rep. 183 (1989) 193.
\bibitem{langa}
P. Langacker, Rev. Mod. Phys. 81 (2009) 1199.
\bibitem{atlasll}
ATLAS Collaboration, arXiv:1707.02424 [hep-ex].
\bibitem{cmsll}
  CMS Collaboration, CMS-PAS-EXO-16-031.
\bibitem{atlasjj}
ATLAS Collaboration, Phys. Rev. D96 (2017) 052004.
\bibitem{cmsjj}
  CMS Collaboration, CMS-PAS-EXO-16-056.
\bibitem{corgen}
G. Corcella and S. Gentile, Nucl. Phys. B866 (2013) 293; 
Erratum-ibid. B868 (2013) 554.
\bibitem{cor}
G. Corcella, Eur. Phys. J. C75 (2015) 264.
\bibitem{gherghetta}
T. Gherghetta, T.A. Kaeding, and G.L. Kane, Phys. Rev. D57 (1998) 3178.
\bibitem{sarah}
F. Staub, Comput. Phys. Commun. 184 (2013) 1792.
\bibitem{spheno}
  W. Porod and F. Staub, Comput. Phys. Commun. 183 (2012) 2458.
\bibitem{madgraph}
J. Alwall et al., JHEP 1407 (2014) 079.
\bibitem{herwig}
  G. Corcella et al., JHEP 0101 (2001) 010.
\bibitem{araz}
  J.Y. Araz, G. Corcella, M. Frank and B. Fuks,  arXiv:1711.06302 [hep-ph].
\end{thebibliography}
\end{document}